\documentclass[FBSedit,FBSmath,ecsub]{FBSsuppl}
\usepackage{amsfonts}
\usepackage{amssymb}
\usepackage{epsfig}

\def\Journal#1#2#3#4 {{#1} {\bf #2}, #3 (#4)}
\def\PR{ Phys. Rev.}
\def\PRC{{ Phys.\ Rev.}\ {\bf  C}}
\def\PRD{{ Phys.\ Rev.}\ {\bf  D}}
\def\PRL{ Phys.\ Rev.\ Lett.}
\def\EPJA{{ Eur.\ Phys.\ J}\ {\bf A}} 
\def\RMP{ Rev.\ Mod.\ Phys.}
\def\ANP {Advances in Nucl.\ Phys.}
\def\PLB{{Phys.\ Lett.}\ {\bf B}}

\def\bea{\begin{eqnarray}}
\def\eea{\end{eqnarray}}

\begin{document}

\title{Relativistic Theory of Few-Body Systems}
\author{Franz Gross\thanks{\textit{E-mail address:} 
gross@jlab.org}}
\institute{Theory Group, MS12H2, Jefferson Laboratory, Newport News, VA 
23606. USA}


\tighten
\maketitle

\begin{abstract}
Very significant advances have been made in the relativistic theory of few
body systems since I visited Peter Sauer and his group in Hannover in
1983.  This talk provides an opportunity to review the progress in this
field since then. Different methods for the realtivistic calculation of few
nucleon systems are briefly described.  As an example, seven relativistic
calculations of the deuteron elastic structure functions, $A$,
$B$, and $T_{20}$, are compared.  The covariant
\textsc{spectator$^{\scriptsize\copyright}$} theory, among the more
successful and complete of these methods, is described in more detail.

\end{abstract}

{\small\vspace*{-4.8in}\rightline{\parbox{1.5in}
                {\leftline{JLAB-THY-02-61}
                \leftline{WM-02-112}
\vspace*{4.5in}}}}
 
\section{Prologue}

In 1983, Peter Sauer invited me to Hannover to lecture to his students on
my \textsc{spectator$^{\scriptsize\copyright}$} approach to the
relativistic theory of few body systems.\footnote{At this conference some
speakers used the term ``spectator'' to refer to an
approximation in point-form quantum mechanics.  When I pointed out that
this usage would increase the confusion already present in this field,
they jokingly suggested that I should copyright the term, and my notation
is a response to this suggestion.} 
This was an exciting time, and I am very glad to have this
opportunity today to thank Peter and his students for their interest, and
for the many good questions and interesting discussions we had during and
after those lectures.  Peter and I never wrote a paper together, but my
collaboration with Alfred Stadler and Teresa Pe\~na eventually grew from
this beginning.   The calculation of the relativistic three nucleon
bound state done with Alfred Stadler is based in part on notes I originally
prepared for the Hannover lectures.  

In preparing this talk I decided to summarize progress made since 1983, as
if I were resuming discussion with Peter again on a Monday morning in 1983
(after an unusually long ``19 year weekend'').  Many of the results
discussed here are covered in more detail in the recent review I wrote
with Ron Gilman \cite{GG}.  

\section{Relativistic approaches reviewed and compared}
\label{sec1}

Since 1983 several alternative relativistic approaches have been
developed.  Attempts to classify and compare them is a continuing
challenge.

These relativistic approaches fall into two major
``schools.''  Both of these schools, and several alternatives within each,
have recently been used to calculate the deuteron structure functions and
form factors relativistically.  These seven calculations (referred to as
the ``c7'' below) are summarized in Table \ref{tab:schools}.  

\begin{table}[t]
\caption{Relativistic calculations referred to in figures~\ref{fig:abt20}
and \ref{fig:F3}. }
\label{tab:schools}
\begin{tabular}{|l|p{70mm}|}
\hline
\tabstrut Greens function dynamics &  \\
\hline
VODG & Manifestly covariant \textsc{spectator$^{\scriptsize\copyright}$}
theory \cite{vdg95} \\
PWD & ``Equal time'' calculation based on the Mandelsweig-Wallace
equation \cite{PW}\\
\hline
\tabstrut Hamiltonian dynamics & \\
\hline
SPR & instant-form; no $v/c$ expansion \cite{FS01}\\ 
ARW & instant-form; with $v/c$ expansion \cite{ARW00}\\
CK & front-form; dynamical light-front \cite{CK99}\\ 
LPS & front-form; fixed light-front \cite{LPS00} \\
AKP & point-form \cite{AKP01}\\ 
\hline
\end{tabular}
\end{table} 

\subsection{Greens function dynamics}
In applications to the deuteron and $NN$ scattering below the pion
production threshold, methods based on Greens function dynamics

\begin{itemize}

\item start from a covariant effective
field theory that takes nucleons and
mesons to be the effective degrees of freedom, 

\item use this effective theory to define the covariant generalized
meson-exchange ladder sum (the sum of all ladder and crossed ladder
diagrams, either omitting vertex corrections and self energies entirely, or
treating them phenomonologically using form factors),

\item  {\it define\/} a free $NN$ progagator, $G_x$, and use it to
reorganize the generalized ladder sum into an infinite series of
``irreducable'' kernels (potentials) $V^{(i)}$ (involving the exchange of
$i$ mesons), and products of these kernels, such as
$V^{(i)}G_xV^{(j)}$, $V^{(i)}G_xV^{(j)}G_xV^{(k)}$, etc.,  

\item evaluate the ladder (ie. one boson exchange) amplitude using
the integral equation $M_x^{{\rm OBE}}=V^{(1)}+V^{(1)}G_xM_x^{{\rm OBE}}$,
and 

\item evaluate the contributions form multi-boson exchange kernels
($V^{(i)}$ with $j\ge2$), perturbatively.

\end{itemize}  
For $j\ge2$, it is usually either assumed that $V^{(j)}\ll V^{(1)}$, or
that 
$\sum_j V^{(j)}\simeq V'^{(1)}$ so the last step need not be done. 
There are several different choices
for the free Greens function, $G_x$, and hence my use of the term ``Greens
function dynamics.''

This approach is manifestly covariant.  This means that the action of {\it
all\/} of the 10 generators of the Poincar\'e group on matrix elements can
be fully specified in terms of the kinematics, so that the action of finite
Poincar\'e transformations can be calculated in simple, closed form. 
Momentum, energy, and angular momentum are strictly conserved, and
finite boosts are given by a simple operation on the matrix elements, with
no further calculations needed.  [In practice, approximations are sometimes
made that spoil this covariance.]  A further advantage is that the effective
field theory used to constuct the generalized ladder sum can also be used
to construct consistent, manifestly covariant electromagnetic currents.  A
very general method of constructing these currents
consistently was published in 1987
\cite{GR87}, opening up the application of these methods to
electromagnetic interactions.  Unfortunately, the construction is not
unique, so additional phenomenology is needed to fix the currents.
 
The disadvantages of the Greens function approach are that  

\begin{itemize}

\item the Greens function, $G_x$, will include the propagation of negative
energy (or antiparticle) states, requiring that the Hilbert space of quantum
mechanical states be expanded, and that we learn to deal with the
mathematical and physical interpretation of these states, 

\item the kernels and the propagator can include unphysical singularities
that must be removed or avoided, complicating either the theoretical
development or the numerical calculations, and

\item the $NN$ scattering amplitudes, which now include contributions from
virtual negative energy states, should be fit directly to the
$NN$ data, thus increaseing the ``investment'' required to develop and apply
these methods.  

\end{itemize}    

\subsection{Hamiltonian dynamics}

In elementary quantum mechanics, states are defined at $t=0$ and their
evolution away from $t=0$ is given by the Schr\"odinger equation
\bea
i\frac{\partial }{\partial t}\Phi({\bf r},t)\Big|_{t=0}=H 
\;\Phi({\bf r},0)\, .
\eea 
In 1949, Dirac \cite{Dirac} pointed out that it is possible to gereralize
the initial hypersurface on which the states are defined, leading also to a
generalization of the Hamiltonian. Methods using Hamiltonian
dynamics are so named because they use one of the three forms of generalized
dynamics identified by Dirac \cite{KP91}.  These
are\footnote{Please beware of some misprints in the discussion of
Hamiltonian dynamics in Ref.
\cite{GG}.}  

\begin{itemize}

\item {\it Instant form\/}: the states are defined on the 
$t=0$ hyperplane, and the Hamiltonian is the usual generator of time
translations, $H$.  The generators of space translations (the momentum
operators $P^i$) and rotations (the angular momentum operators $J^i$) form
a subalgebra of the Poincar\'e group that leaves the $t=0$ hypersurface
invariant, so they may be defined without regard to the interactions living
in $H$.  The generators of the boosts, $K^i$, do not leave the hyperplane
invariant, so must also include the interactions, and this is the
disadvantage of this method.

\item {\it Front form\/}: the states are defined initially on the
light-front $t_+\equiv t+z=0$, and the generalized Hamiltonian is
$H_-\equiv H-P^z$.  The subgroup leaving this light-front invariant is
composed of the {\it seven\/} generators $H_+$, $K^z$,
$J^z$, ${\bf E}_\perp=\{K^x+J^y,K^y-J^x\}$, and
${\bf P}_\perp=\{P^x,P^y\}$.  
The dynamical interactions are in the three gererators $H_-$, and
${\bf J}_\perp\simeq\{J^x,J^y\}$.  This method is very popular for high
energy interactions where the dynamics evolves along the light cone $t=z$
and it is a great advantage to have the boost $K^z$ among the kinematic
generators.  Its disadvantage is that the conservation of angular momentum
becomes a dynamical issue.

\item {\it Point form\/}: Here the states are defined on the forward
hyperboloid:  $t>0$ and $t^2-{\bf r}^2=a^2$.  The homogeneous Lorentz group,
generated by the boost and angular momentum operators, $K^i$ and $J^i$,
leaves this surface invariant, and the dynamics are in $H$ and the momentum
operators $P^i$.  This method automatically conserves angular momentum and 
gives boosts independent of the dynamics, but its disadvantage is
that the conservation of momentum becomes a dynamical issue.      

\end{itemize}
All of these methods use quantum mechanical states that span the
positive energy spectrum only, leading to the standard interpretation of
quantum mechanics. In each case it is possible to transform the generalized
Schr\"odinger equation into an equivalent instant form, and hence use the
excellent phenomenological potentials that have recently been fit to low
energy $NN$ scattering.

\begin{figure}[t]
\begin{center}
\vspace*{-0.3in}
\mbox{
   \epsfysize=5.0in
\epsffile{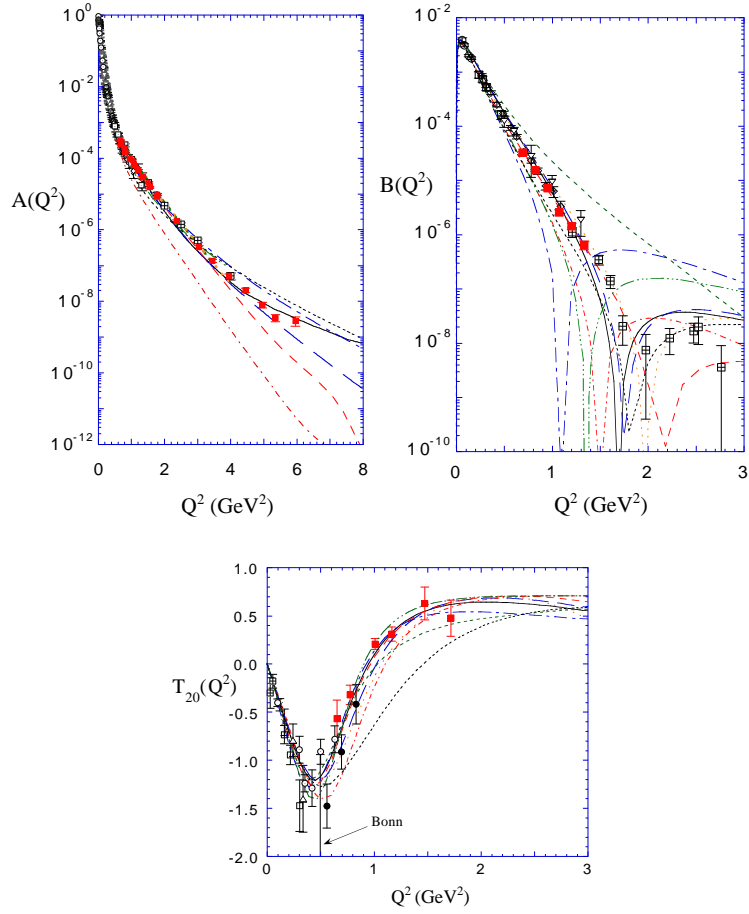} 
} 
\end{center}
\vspace*{-0.1in}
\caption{Predictions for the structure functions $A$, $B$, and
$T_{20}$.   The models, in order of the $Q^2$ of their mimima in $B$, are:
CK (long dot-dashed line), PWD (dashed double-dotted line),
AKP (short dot-dashed line),
VODG full calculation (solid line), VODG in RIA (long dashed line), LPS
(dotted line), a quark model calculation not discussed here (widely spaced
dotted line), SPR (medium dashed line), and ARW (short dashed line).  None
of the curves shown include the $\rho\pi\gamma$ exchange current.  (This
figure is taken from Ref. \cite{GG}.)  }
\label{fig:abt20}
\end{figure}

The disadvantages of Hamiltonian dynamics depend in part on the specific
form chosen.   In the instant form it is difficult to boost the states, and
the front and point forms require special care to insure that either
angular or linear momentum is conserved.  Perhaps the most serious
disadvantage of Hamiltonian dynamics is that they provide litte or no
guidance on how to construct consistent, conserved, covariant
electromagnetic currents.  The construction of currents is
phenomenolgical and sometimes ad-hoc.

\subsection{Predictions for the deuteron form factors}

The predictions of the c7 listed in Table \ref{tab:schools} are
shown in figure \ref{fig:abt20}.  I conclude the following:
\begin{itemize}

\item Only the VODG and SPR calculations provide a reasonable description of
{\it all three\/} of these observables.  This is probably due to the fact
that these are the only calculations valid to all
orders in $(v/c)^2$ that also include complete
currents consistent with the dynamics.

\item The magnetic structure function, $B$, is by far the most sensitive
to the calculations, and provides a stringent test of the theory. 

\end{itemize}

To get some insight into the differences between these calculations, it is
useful to think of each as built from the same three (roughly defined and
overlapping) ingredients:  (i) the ``leading''
nonrelativistic $S$ and $D$-state deuteron wave functions, (ii) the
single nucleon electromagnetic current, and (iii) all the rest, including
$I=0$ interaction currents and relativistic modifications (which are model
dependent and not small) to both wave functions and currents.  

All of the c7 use modern
``realistic'' deuteron wave functions, so the differences between them is
not due to the choice of the leading, nonrelativistic part of the wave
function.  The large differences arise instead from the treatment of
relativistic effects, and the construction of the current. 

Some of the c7 include relativistic effects to all orders in
$(v/c)^2$, while others only include the lowest order terms.  By expanding
in powers of $(v/c)^2$ it would be possible to compare the relativisitc
effects obtained from all of the c7.  To a limited extent this has already
been done; in the 1970's and 1980's Friar and others found that the one pion
exchange corrections for the instant-form and
\textsc{spectator$^{\scriptsize\copyright}$} theory  agreed to lowest
order (for a review see Ref. \cite{Gr91}).  The comparison was very
informative -- it showed that the ``pair term'' corrections
(for example) differed in the two cases, and agreement was only obtained
after {\it all\/} of the corrections were added together.  This leads to
the expectation that a similar comparison of the c7 would lead to
agreement {\it only for those calculations that include all possible
corrections\/}.  I believe that such a benchmark calculation comparing the
different methods could significantly increase our understanding.

Another difference in the c7 is the choice of current.  Even the one body
current is not the same for all of the c7.  For example, in the
point-form it turns out \cite{AKP01} that the single nucleon form factors
must be evaluated at a momentum transfer $Q^2$ {\it larger\/} than that
transmitted to the deuteron as a whole (possible because momentum
is not conserved).  Because the nucleon form factors decrease rapidly in
$Q^2$, this leads to a strong suppression of the prediction for
$A$, as shown in figure~\ref{fig:abt20}.  This could be corrected by adding
an interaction current, but the point-form approach itself does not tell us
what this should be. In the absence of an underlying field theory to guide
the physics, the currents are purely phenomenological.

The currents in the \textsc{spectator$^{\scriptsize\copyright}$} theory are
well constrained, but even here they cannot be uniquely
determined.  I will conclude this discussion by showing how we can exploit
this flexibility, and adjust the single nucleon 
\textsc{spectator$^{\scriptsize\copyright}$} current to give an excellent
fit to all of the elastic electron-deuteron data.  

Following Ref. \cite{GR87}, current conservation requires that the single
nucleon current satisfy the Ward-Takashi identity
\bea
q_\mu \,j_N^\mu(p',p)= S^{-1}(p)-S^{-1}(p') \label{eq2}
\eea     
where $S(p)$ is the propagator of a nucleon with four-momentum $p$.  The
$NN$ model used in the VODG calculation \cite{GVH92} uses the dressed
nucleon propagator  
\bea
S(p)=\frac{h^2(p)}{m-\not\!p}=\frac{h^2(p)}{\Lambda_-(p)}
\eea     
with $h(p)$ a phenomenological scalar function of $p^2$.  The {\it
simplest\/} solution of Eq. (\ref{eq2}) gives the following one nucleon
current 
\bea
j_N^\mu(p',p)=
F_0\left\{F_1(Q^2)\gamma^\mu 
+F_2(Q^2)\frac{i\sigma^{\mu\nu}\,q_\nu}{2m}\right\}
+G_0F_3(Q^2) \Lambda_-(p')\gamma^\mu \Lambda_-(p)
\eea     
where terms proportional to $q^\mu$ have been dropped (they are required by
the identity (\ref{eq2}) but vanish when contracted into the conserved
electron current), 
\bea
F_0&=&\frac{h(p)(m^2-p'^2)}{h(p')(p^2-p'^2)} -
\frac{h(p')(m^2-p^2)}{h(p)(p^2-p'^2)} \nonumber\\
G_0&=&\left[\frac{h(p')}{h(p)} - \frac{h(p)}{h(p')}\right]
\frac{4m^2}{p^2-p'^2}\, ,
\eea     
and $F_3(Q^2)$ is {\it completely undetermined\/} except for the requirement
that $F_3(0)=1$.  The VODG calculation shown in figure \ref{fig:abt20} uses
the ``standard'' dipole form for $F_3$ 
\bea
F_3(Q^2)=\left(\frac{\Lambda^2}{\Lambda^2-Q^2}\right)^2 \label{eq:dipole}
\eea
with $\Lambda^2=0.71$ GeV$^2$.  Clearly other choices of $F_3$ are
possible. 

Another source of uncertainty is the famous $\rho\pi\gamma$
exchange current.  This current is separately gauge invariant and strongly
dependent on the $\rho\pi\gamma$ form factor,
$f_{\rho\pi\gamma}(Q^2)$ \cite{Ito93}.  The value of
$f_{\rho\pi\gamma}(0)$ is constrained by meson radiative decays, and its
contribution to the deuteron form factors at small $Q^2$ is
negligible.  Its importance at high $Q^2$ depends strongly on the assmued
dependence of $f_{\rho\pi\gamma}(Q^2)$, which is unknown but
can be estimated from quark models.  One of the best calculations of this
form factor is that of the Rome group \cite{Rome2}. To clarify the
comparison between models, the versions of the c7 shown in figure
\ref{fig:abt20} did not included this current, and it could be added to any
of them.

\begin{figure} 
\begin{center}
\vspace*{-0.2in}
\mbox{
   \epsfysize=6.5in
\epsffile{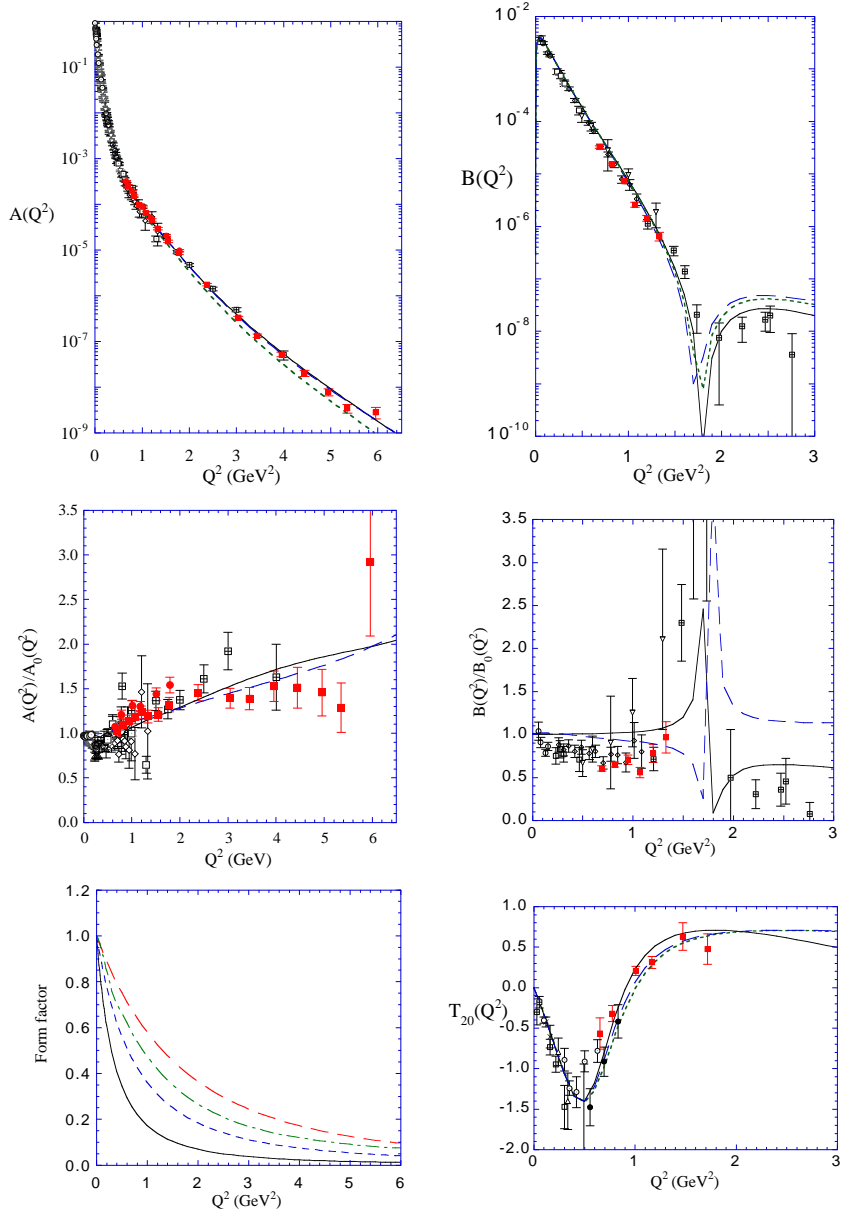} 
} 
\end{center}
\vspace*{-0.1in}
\caption{Upper panels ($A$ and $B$) and lower right panel
($\tilde T_{20}$) compare data to three theoretical
models based on VODG: (i) ``standard'' case referred to in
the text (dotted line), (ii) model with the tripole $F_3$
(solid line), and (iii) model with the dipole
$f_{\rho\pi\gamma}$ (dashed line).  The center two panels
show the data and models (ii) and (iii) divided by model
(i).  The lower left panel shows form factors: standard dipole
with $\Lambda^2=0.71$ (solid line), dipole with
$\Lambda^2=1.5$ (short dashed line), Rome
$f_{\rho\pi\gamma}$ (dot-dashed line) \cite{Rome2}, and the
tripole with $\Lambda^2=5$ (long dashed line). (This
figure is taken from Ref. \cite{GG}.)}
\label{fig:F3}
\end{figure} 

Taking the  VODG calculation in RIA approximation (discussed in Ref.
\cite{vdg95}) as our ``standard,'' I now consider the effect of (i)
altering the $Q^2$ dependence of $F_3$, or (ii) adding the
$\rho\pi\gamma$ exchange current.  The ``standard'' calculation is the
long dashed lines shown in figure \ref{fig:abt20} and the dotted lines
reproduced in figure \ref{fig:F3}.  The effect of changing $F_3$ to a
tripole form
\bea
F_3(Q^2)=\left(\frac{\Lambda^2}{\Lambda^2-Q^2}\right)^3
\eea
with $\Lambda^2=5$ GeV$^2$ (and continuing to keep $f_{\rho\pi\gamma}=0$)
are shown as solid lines.  The effect of adding a $\rho\pi\gamma$
exchange current using a dipole form factor with $\Lambda^2=1.5$ GeV$^2$
(and keeping the standard dipole form for $F_3$) are shown as the dashed
curves.  [The $F_3$ and $\rho\pi\gamma$ form factors themselves are
compared to the Rome form factor and the standard dipole (\ref{eq:dipole})
in the lower left panel.]  To see the effects on
$A$ and $B$ more clearly, the middle two panels show the ratios of $A/A_0$
and $B/B_0$, where $A_0$ and $B_0$ are the standard calculation.

We see that a reasonable adjustment of $F_3$ at high $Q^2$ can give an
excellent description of all of the elastic deuteron observables.  This is
perfectly permisable within the theory, since the form factor $F_3$, while
it {\it must\/} be present, cannot be determined by on-shell data, and must
be treated phenomenologically.  From this point of view the deuteron
data has now determined the unknown form factor $F_3$, and the isoscalar
single nucleon current is now fixed.  It remains to be seen whether
the same $F_3$ will give excellent results for other electron
scattering observables.

What are we to say about the $\rho\pi\gamma$ exchange current?  My own
belief is that this exchange current is being overestimated, even
using the Rome form factor.  While this current is certanly present, it
is probably either negligible, or cancelled by the many other short range
currents neglected in these calculations.  

\section{Progress with the {\sc{spectator}}$^{\copyright}$ theory} 

In the remainder of this talk I return to the
\textsc{spectator$^{\scriptsize\copyright}$} theory and report on
developments since 1983, and on my expectations for the future.

\subsection{Definition of the theory}

The \textsc{spectator$^{\scriptsize\copyright}$} propagator for two
nucleons with four momenta $p_1$ and $p_2$ is 
\bea
G_S= 2\pi i\,\delta_+(m^2-p_2^2)\,S(p_1)\, 
\sum_\lambda u({\bf p}_2,\lambda)\bar u({\bf p}_2,\lambda)
\eea
where the $\delta_+$ function restricts $p_{20}$ to its positive energy
mass-shell, so that $p_2=\{E(p_2),{\bf p}_2\}$, and $u({\bf p}_2,\lambda)$
is the free Dirac spinor for a nucleon with three-momentum ${\bf
p}_2$ and helicity $\lambda$, normalized to $\bar u({\bf p},\lambda) u({\bf
p},\lambda') =2m\delta_{\lambda \lambda'}$.  This propagtor insures that
the integration over the internal energy will place particle 2 on its
positive energy mass-shell, giving the 
\textsc{spectator$^{\scriptsize\copyright}$} equation.  It was already known
before 1983 that this equation is manifestly covariant, has the right
one-body limit, a smooth nonrelativistic limit, satisfies the cluster
property, and produces a gauge invariant one-photon-exchange interaction
\cite{Grosseq}.  The manifest covariance means that momentum and angular
momentum are trivially conserved, and that the boosts are known exactly, to
all orders in
$(v/c)^2$.  

\subsection{Progress since 1983} 

Progress since 1983 has been substantial.  Theoretically, we
have seen (i) the development of a general method for
constructing consistent, conserved currents \cite{GR87} (already discussed),
and (ii) the recent demonstration that all vertex corrections and momentum
dependent self energies {\it cancel\/} in massive scalar QED \cite{SGT02}. 
This last result means that the generalized ladder sum gives the full
result for the interaction of scalar bosons with a massive photon. 
Combining this with the already known fact that the generalized ladder sum
is well approximated by the ladder approximation to the
\textsc{spectator$^{\scriptsize\copyright}$} equation, and we have the first
demonstration that solutions of the
\textsc{spectator$^{\scriptsize\copyright}$} equation are a good
approximation to the {\it exact\/} solutions of massive scalar QED.  It is
not known if this result also holds for other effective theories.

Many applications have been developed since 1983.  These include:

\begin{itemize}

\item Mesons as $q\bar q$ bound states, including covariant
treatment of scalar confinement consistent with chiral symmetry breaking
\cite{GMall,SG01}.

\item Unitary model of $\pi N\to \pi N$ and $\gamma N\to \pi N$ processes up
to 600 MeV including $\Delta$, Roper, and $D_{13}$ resonances.
\cite{GSall}.  

\item $NN$ scattering and bound state solutions to 350 MeV lab kinetic
energy with $\chi^2\sim2$ and 13 fitting parameters \cite{GVH92}. 

\item Exact numerical solution of the relativistic three body equations
that give a good binding energy for $^3H$  without relativistic three
body forces \cite{SG97}.

\item Good description of the deuteron form factors \cite{vdg95}
reviewed above.

\item General classification of inelastic scattering observables \cite{DG89}
and first calculation of inelastic scattering in relativistic impulse
approximation \cite{edinelastic02}.

\item Derivation of the $pA$ multiple scattering series \cite{GM90} and
applications to proton nucleus scattering \cite{Metal}. 

\end{itemize}

When I visited Peter Sauer in 1983, the goal was to do a relativistic 
three-body calculation.  Peter and I never completed this, but my
contacts with Peter lead eventually to my collaboration with Alfred
Stadler, and to the desired calculation.  

\begin{figure} 
\begin{center}
\vspace*{-0.2in}
\mbox{
   \epsfxsize=3.0in
\epsffile{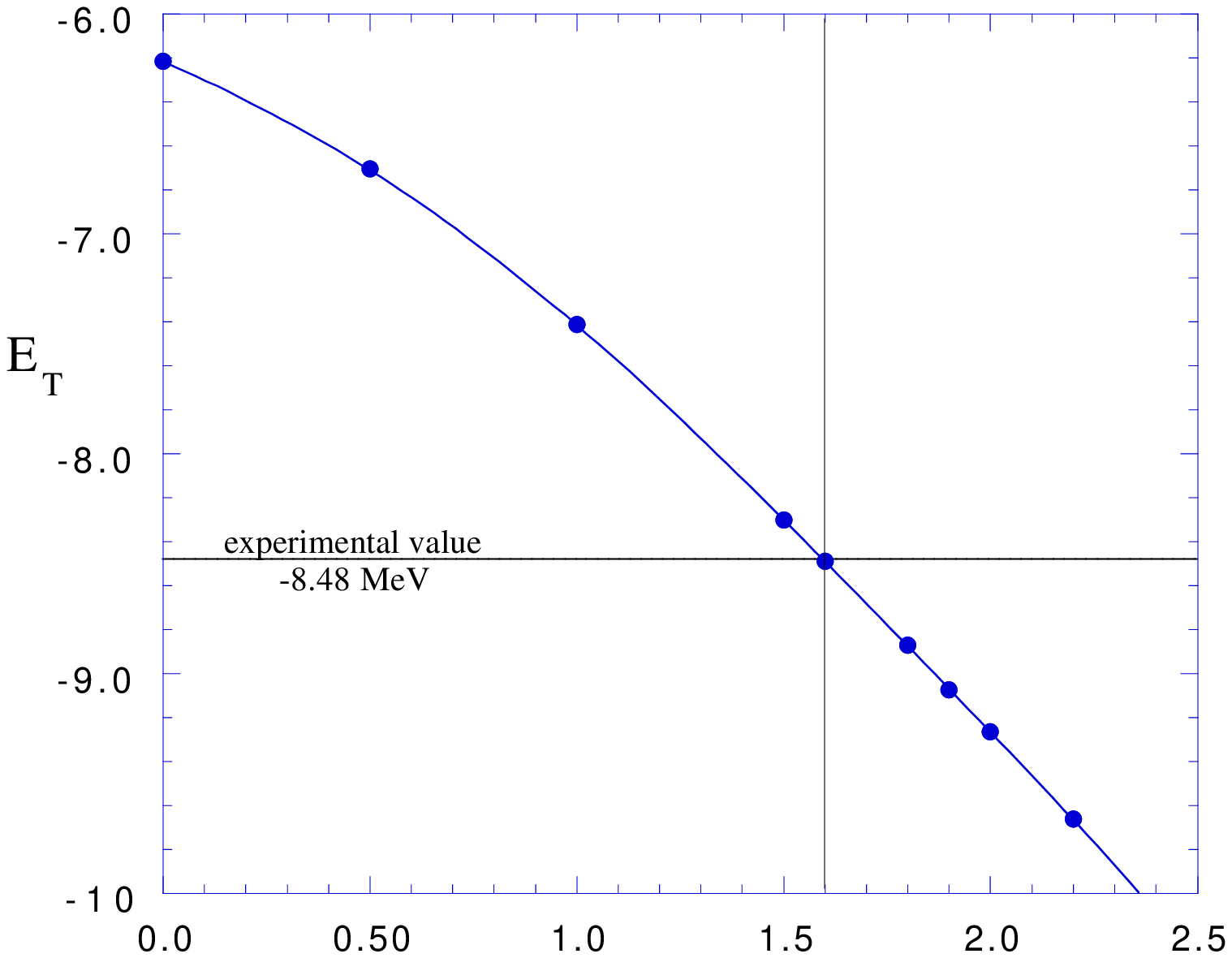}} 
\mbox{
   \epsfxsize=3.0in
\epsffile{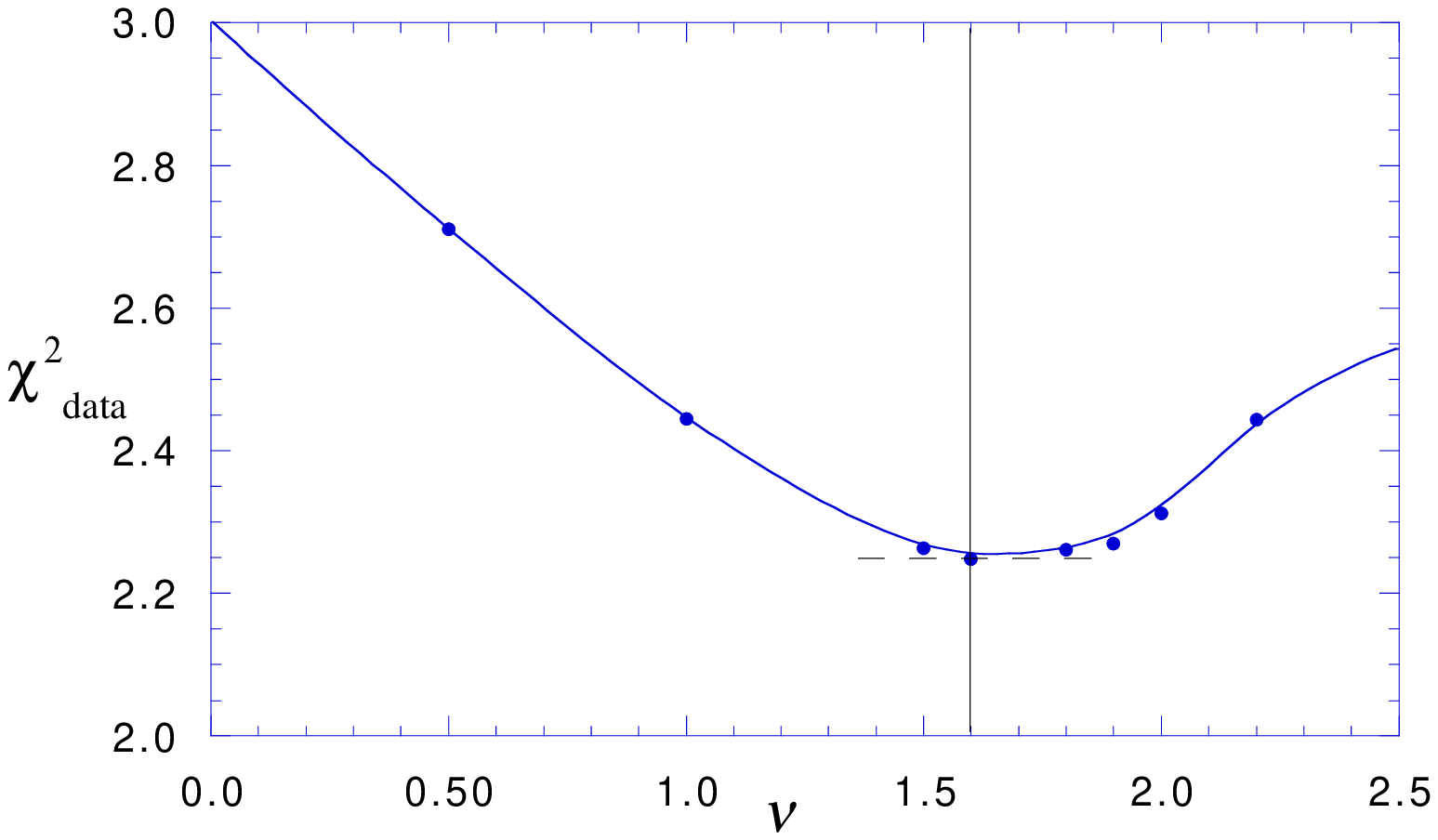} 
} 
\end{center}
\vspace*{-0.1in}
\caption{Upper panel shows the three body binding energy as a function
of the off-shell coupling parameter $\nu$ discussed in the text.  The
lower panel shows the variation in $\chi^2$ of the fit to the two-body
data (up to lab energies of 350 MeV).}
\label{fig:3body}
\end{figure}

The major result of this relativistic calculation of the three body binding
energy \cite{SG97} is shown in figure \ref{fig:3body}.  We obtain good
agreement with the binding energy only after introducing
an off-shell scalar meson coupling
\bea
g_s\Lambda(p',p)=g_s\left\{1 
-\frac{\nu_s}{2m}\left[\Lambda_-(p')+\Lambda_-(p)\right]\right\}
\eea
with the off-shell coupling parameters $\nu_\sigma$ and $\nu_\delta$
proportional to the $\nu$ shown in figure \ref{fig:3body}.  The exciting
conclusion is that the same $\nu$ that fits the experimental binding energy
also gives the best fit to the two-body data.

\subsection{The NEW \textsc{spectator$^{\scriptsize\copyright}$} theory}

The \textsc{spectator$^{\scriptsize\copyright}$} theory raises a number of
questions:

\begin{figure} 
\begin{center}
\vspace*{-0.1in}
\mbox{
 \epsfxsize=5.0in
\epsffile{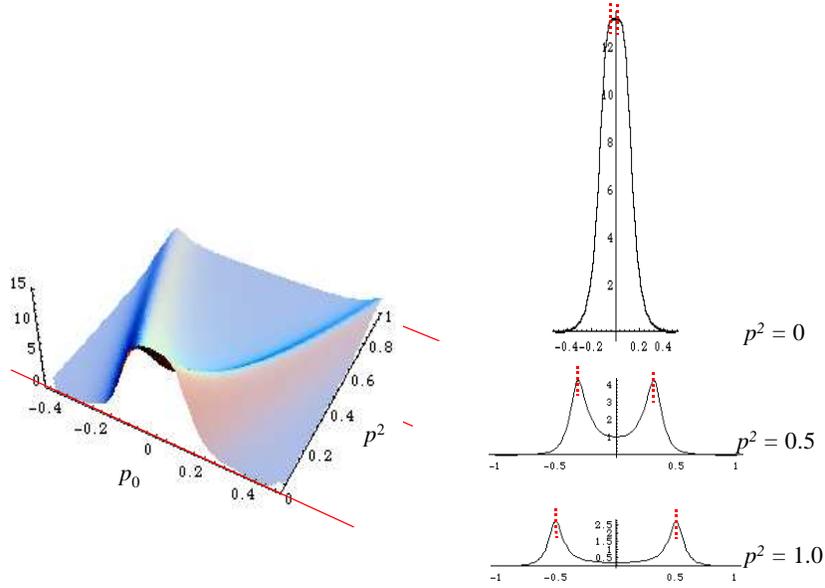} 
} 
\end{center}
\vspace*{-0.5in}
\caption{Mass-shell peaks in the free propagator for two particles
with equal mass $m$ a total rest mass of $M/m=1.8$.  The figures on the
right show slices through the surface at various constant values of $p^2$. 
(All quantities are in units of
$m$.)}
\label{fig:prop}
\end{figure} 

\begin{itemize}
\item In the $n$-body problem, why should $n-1$ particles be on shell,
and which ones should they be?

\item Can all spurious singularities be removed from the theory?

\item What are the Feynman rules for the general case?

\item Can the theory be formulated as a quantum mechanics, with a well
defined Hilbert space of states? 
 
\end{itemize}
Many of these questions were raised initially by Peter Sauer and his
students during my visit in 1983, and I have answered some of them to my
satisfaction.  I am currently preparing a new formulation of the 
\textsc{spectator$^{\scriptsize\copyright}$} theory that I believe will
provide more complete answers.  Like a good
salesman of computer games, I want to advertize my product in advance of
its ``public'' release.

Figure \ref{fig:prop} shows the real part of Bethe-Salpeter propagator,
$iG_{BS}$, as a function of the relative energy
$p_0=\frac{1}{2}(p_{10}-p_{20})$ and the square of the relative
three-momentum, $p^2$, evaluated in the c.m. system (to aid in seeing the
structure the $\epsilon$'s in the $-i\epsilon$ prescriptions were given a
finite value).  This figure shows clearly that the propagator is
dominated by the two positive energy mass-shell poles.  Using Cauchy's
residue theorem to evaluate the integral over $p_0$ gives the result that
the integration is dominated by the on-shell contribution from {\it one\/}
particle, but when interpreted as an ordinary integral along the real axis
the same result requires the contribution of {\it both\/} peaks (each giving
one-half of the residue result).  Looked at from this point of
view, the wave function is the sum of two terms, one with
particle 1 on-shell and one with 2 on-shell.  These two terms have support
in separate regions of phase space, so adding them together leaves some
things unchanged, but continuing in this direction leads to a
reformulation of the \textsc{spectator$^{\scriptsize\copyright}$} theory
that gives nice answers to the questions above.  Watch for the release of
this theory in the near future!

\begin{acknowledge}
I thank Peter Sauer for stimulating so many 
developments in Few Body physics, Alfred Stadler for his contributions
to the relativistic three-body problem,	and for conversations leading to
the new \textsc{spectator$^{\scriptsize\copyright}$} theory, and John Tjon,
Wally Van Orden, and Teresa Pe\~na for many useful discussions and long
and fruitful collaborations.  This work was supported in part by DOE
grant No.~DE-FG02-97ER41032, held at William and Mary.  The Southeastern 
Universities Research Association (SURA) operates the Thomas Jefferson
National Accelerator Facility under DOE contract DE-AC05-84ER40150.

\end{acknowledge}

\end{document}